\documentclass[preprint,12pt,authoryear]{elsarticle}

\usepackage{graphicx}
\usepackage{subfigure}
\usepackage[fleqn]{amsmath}  
\usepackage{amsfonts}
\usepackage{amssymb} 
\usepackage{amsbsy}
\usepackage{lineno}
\usepackage{mathrsfs}
\usepackage{times}
\usepackage{multicol}
\usepackage{multirow}
\usepackage{geometry}
\usepackage{caption}
\usepackage{array}
\usepackage{float}
\usepackage{rotating}
\usepackage{booktabs}
\usepackage{fancyhdr}
\usepackage{cases}
\usepackage{epstopdf}
\usepackage{blindtext}
\usepackage{dblfloatfix} 
\usepackage[export]{adjustbox}
\usepackage{romanbar}
\usepackage[autostyle=false, style=english]{csquotes}
\MakeOuterQuote{"}
\usepackage[pdftex,bookmarks,colorlinks=true,linkcolor=blue,citecolor=blue]{hyperref}

\begin{document}
\begin{frontmatter}

\title{Empirical study of effect of dynamic travel time information on driver route choice behavior}

\author{Jinghui Wang\fnref{label1}}
\author{Hesham A. Rakha\corref{cor}\fnref{label2}}
\cortext[cor]{Corresponding Author.}
\ead{hrakha@vt.edu}
\address[label1]{Center for Sustainable Mobility,Virginia Tech Transportation Institute,3500 Transportation Research Drive,Blacksburg,VA 24061,USA}
\address[label2]{Center for Sustainable Mobility,Virginia Tech Transportation Institute,3500 Transportation Research Drive,Blacksburg,VA 24061,USA}
\begin{abstract}
The objective of this paper is to study the effect of travel time information on day-to-day driver route choice behavior. A real-world experimental study is designed to have participants repeatedly choose between two alternative routes for five origin-destination pairs over multiple days after providing them with dynamically updated travel time information (average travel time and travel time variability). The results demonstrate that historical travel time information enhances behavioral rationality by 10\% on average and reduces inertial tendencies to increase risk seeking in the gain domain. Furthermore, expected travel time information is demonstrated to be more effective than travel time variability information in enhancing rational behavior when drivers have limited experiences. After drivers gain sufficient knowledge of routes, however, the difference in behavior associated with the two information types becomes insignificant. The results also demonstrate that, when drivers lack experience, the faster less reliable route is more attractive than the slower more reliable route. However, with cumulative experiences, drivers become more willing to take the more reliable route given that they are reluctant to become risk seekers once experience is gained. Furthermore, the effect of information on driver behavior differs significantly by participant and trip, which is, to a large extent, dependent on personal traits and trip characteristics. 
\end{abstract}

\end{frontmatter}
%
\section{Introduction}
%
%
%
%
Advanced traveler information systems (ATISs), which are an integral component of Intelligent Transportation Systems (ITSs), are designed to provide real-time information that enables drivers to choose rationally from among alternative routes. The effectiveness of ATIS is dependent on drivers response to received information. Incorporating information into the modeling practice may enhance the accuracy of route choice models by adding realistic behavioral mechanisms and thus improve the effectiveness of ITSs. Accordingly, it is essential to capture the behavioral generalization of informed drivers in order to enhance ATIS design.\par

From a modeling perspective, traditional transportation research attempts to replicate driver route choice behavior assuming that individuals are capable of accurately perceiving route performance and attempt to maximize their expected utility. Mathematically, such assumptions make it cost-effective and technically simpler to model traveler behavior. Most attempts at route choice modeling are discrete choice models that are econometrically derived from random utility theory. Since the 1970\textquoteright s, transportation researchers have studied the decisions associated with route choice modeling. In the past forty years, innovations in discrete choice models has progressed in three stages, namely: Multinomial Logit Modeling \cite{daganzo1977stochastic}, Nested Logit Modeling \cite{ben1985discrete} and Mixed Logit Modeling (Ben-Akiva et al. unpublished manuscript, 1996). Each enhancement attempted to direct the logit model towards more flexible model structures. Despite previous achievements, the ability of these models to capture realistic route choice behavior has been increasingly challenged due to insights from the psychology field. Through a range of empirical research, drivers were detected to be not omniscient, as expected in traditional models, in precisely perceiving the actual route performance. Bounded rationality was initially introduced by Simon \cite{simon1982models} to explicitly account for the fact that human beings are incapable of identifying the best route among multiple alternatives due to limitations in knowledge, cognition and information acquisition. Tawfik and Rakha \cite{tawfik2012real} verified Simon's theory using a real world experiment, demonstrating that drivers generally only had a 50\% accuracy in perceiving route information (e.g. travel time, travel distance, speed, etc.). Even though travelers occasionally have correct perception of route performance, they may not be willing to switch to the perceived better route; rather, they stick to the habitual choice until its performance is not satisfying. In other words, travelers are not necessarily utility maximizers \cite{vreeswijk2013drivers}. Satisficing psychology triggers individuals' behavioral mechanisms in seeking for a satisfactory solution instead of the optimal one. Irrational behaviors deviate travelers from the best route and are not easily predictable by traditional models due to limitations in model assumptions.\par

Route information provides an explicit description of the actual performance of the choice sets, which has the potential to improve travelers knowledge and direct them towards the objectively optimal decision. Accordingly, route information is expected to facilitate travelers to make more logical choices (choose faster routes in this study). The accuracy of traditional discrete choice models may probably be improved by integrating information effects into the modeling practice. An explicit generalization of the effect of information on route choice behavior is thus studied.\par

The proposed research attempts to provide valuable insights in addressing a number of important questions, namely: does route information enable drivers to behave more rationally? How does the information affect behavioral mechanisms from person to person as well as from trip to trip? What is the difference in behavioral effects between information types? This study is a follow-up experiment of \cite{tawfik2012real}. The results of the two experiments are compared and provide significant implications to the behavioral effect of real-time information. The major contribution of this study is to design a real world route choice experiment and study realistic route choice behavior of informed drivers and their day-to-day behavioral variations. This study differs from most of the studies in the literature that have investigated driver route choice behavior in a hypothetical environment such as simulation and questionnaire that is unable to completely reflect reality. The paper also comprehensively presents the heterogeneity of drivers' responses to the provided route information, considering the diversities in driver's age, gender, and personal traits, trip characteristics, and temporal variation. The findings of this study are critical and insightful to the modeling of route choice behavior and personalized ATIS design. 
\section{Literature Review}
Empirical research found that the factors considered by travelers in making route choice decision were not unitary \cite{vreeswijk2013drivers}. Numerous attributes were found to be important considerations, including travel time, trip distance, average speed, and the number of traffic signals along the route. Nonetheless, previous attempts at identifying the attributions of route choice identify travel time as the most important factor even though travelers may also consider other factors. In accordance with Tawfik and Rakha's study, 70\% of drivers' route choices was successfully explained by travel time followed by average speed and distance traveled \cite{tawfik2010experimental, tawfik2012real}. Consequently, travel time information is provided to the test participants in this study.\par

As captured in the “hot stove” effect \cite{denrell2001adaptation}, individuals were not inclined to select options associated with high variability, although these might actually provide larger benefits. Considering uncertainty, people do not have perfect knowledge of the gains that could be accrued and the loss associated with risking changing habitual choices. Prospect Theory \cite{kahneman1979prospect} explicitly and thoroughly describes this psychological behavior that risk-seeking behavior would likely exhibit in the loss domain rather than in the gain domain. In relation to route choice, katsikopoulos et al. \cite{katsikopoulos2002risk} verified the results of Prospect Theory through a simulated experiment in which participants were provided with the information of travel time variability, indicating that risk aversion emerged in the gain domain (alternative route is faster but riskier) while risk seeking emerged in the loss domain (alternative route is slower but riskier). Accordingly, drivers repeatedly make illogical choices due to the risk aversion in the gain domain. Information is expected to reduce the uncertainty and enhance rational behavior partially by leading travelers to risk seeking in the gain domain. katsikopoulos et al. \cite{katsikopoulos2000framing,katsikopoulos2002risk} revealed that the provided information supported for choice rationality and reduced inertia. \par

The behavioral effect of travel time information on route choice behavior has been incrementally studied both from a theoretical and practical standpoint. Early studies, such as Lida et al. \cite{iida1992experimental} and Yang et al. \cite{yang1993exploration}, pioneered the investigation of the information effects on drivers route choice behavior, both of which conducted studies in the simulation environment. Ben et al. \cite{ben2008combined} thoroughly investigated the combined effects of information and driving experience on route choice behavior using a simulated experiment. The results provided evidence to suggest that the expected benefit of information is achieved only if drivers lacked long-term experience. Based on this study, a discrete choice model with Mixed Logit specifications was developed to accurately describe the respondents' learning process under the provision of real-time information \cite{ben2010road}. Further, Ben et al. \cite{ben2010road} also demonstrated that information provided on average travel time resulted in different responses compared to information on travel time variability, which remains to be verified. Using a simulation- and a stated preference-based approach, numerous attempts were made to econometrically address the various behavioral mechanisms of drivers\textquoteright ~route choice with real-time information. The studied behavioral mechanisms involved logical choice \cite{ben2008combined,ben2010road}, inertia choice \cite{srinivasan2000modeling, katsikopoulos2000framing}, switching behavior \cite{jou2005route,polydoropoulou1994influence,srinivasan2003analyzing}, habit and learning \cite{bogers2005joint,tawfik2013latent}, and others \cite{lee2016valuation, kou2017urban, moghaddam2017effect, moghaddam2019comprehending, dai2019simulation, su2019edge}. Specifically, Karthik et al. \cite{srinivasan2000modeling} demonstrated that user experiences decreased inertia behavior in day-to-day variation. The travel time information was demonstrated by many studies to effectively move route choice towards rationality (\cite{ben2008combined, ben2010road, srinivasan2003analyzing, jou2005route, avineri2006impact, dia2002agent, srinivasan1999role, moghaddam2019comprehending, dai2019simulation}), however, the effect of information strongly depends on other factors, such as personal traits, trip characteristics, and other decision considerations. From the personal trait perspective, Jou et al. \cite{jou2005route} concluded that elderly travelers would be less likely to switch due to the habitual and risk-aversive effects, and male travelers would be more likely to switch to the best route. Also, trip characteristics and traveler preferences were proved by Polydoropoulou et al. \cite{polydoropoulou1994influence} to significantly affect route switching and compliance with information. In summary, to the authors' best of knowledge, existing studies have typically lacked realism (either based on simulation or stated preferences approaches) and have not characterized the effect of information details of trip characteristics, such as directness of the route, number of intersections, conflicts with non-motorized traffic on driver route choice behavior. This study attempts to address this void. \par

Although previous attempts provided econometric and empirical generalizations, most were based on simulation and stated preference approaches. In the simulator surroundings, however, respondents make decisions in a digital and virtual environment. Stated preference is an investigative approach in which respondents are given questionnaires to make choices hypothetically. Both approaches are performed under fictitious conditions and may not accurately capture actual choice behavior. Consequently, an in-field case study is needed to address the driver route choice behavior. To the author's best of knowledge, this study, is the first attempt at addressing this need using dynamic travel time information, which differs from the previous real-world experiments (e.g. \cite{ramaekers2013modelling, papinski2009exploring, li2005analysis, mahmassani2000transferring} that conducted experiments for a short time period (e.g. several days) and did not capture the day-to-day variation of route choice behavior using the learning mechanism that accounts for information effects. As a follow-up test of Tawfik and Rakha's experiment (in which information was not available), participants were provided with real-time information.\par

Drivers' responses to information may differ based on personal characteristics, demographics, preferences and choice situations \cite{abdel1997using,parkany2004attitudes}. Nonetheless, few studies so far have attempted to quantitatively investigate such discrepancy. Tawfik et al. \cite{tawfik2013latent} developed a latent class choice model by classifying personal traits and choice situations into four behavioral groups as illustrated in \autoref{tab.1}. The results demonstrated that the model outperformed traditional hierarchical models in predicting realistic behavior. However, Tawfik et al.'s study did not incorporate the effect of information in the modeling practice. Accordingly, this study attempts to investigate the information effect considering different participants and choice situation characteristics in order to capture preliminary insights for modeling in the future horizon.\par

In general, given the incomplete picture of the behavioral aspects of route-choice decision making, more attempts are justified. The proposed research is thus initiated by a real world case study to provide a better understanding of underlying behavioral effects of travel time information on route choice decisions.\par

\begin{table*}
\centering
\caption{Four identified behavioral driver types \cite{tawfik2013latent}}
\label{tab.1}
\small
\renewcommand{\arraystretch}{1.25}
\begin{tabular}{@{}ccp{6cm} @{}}\toprule[1.5pt]
 Behavior Type& Typical Behavior & Type Description \\\midrule 
  1& \raisebox{-3.0cm}{\includegraphics[scale=0.5]{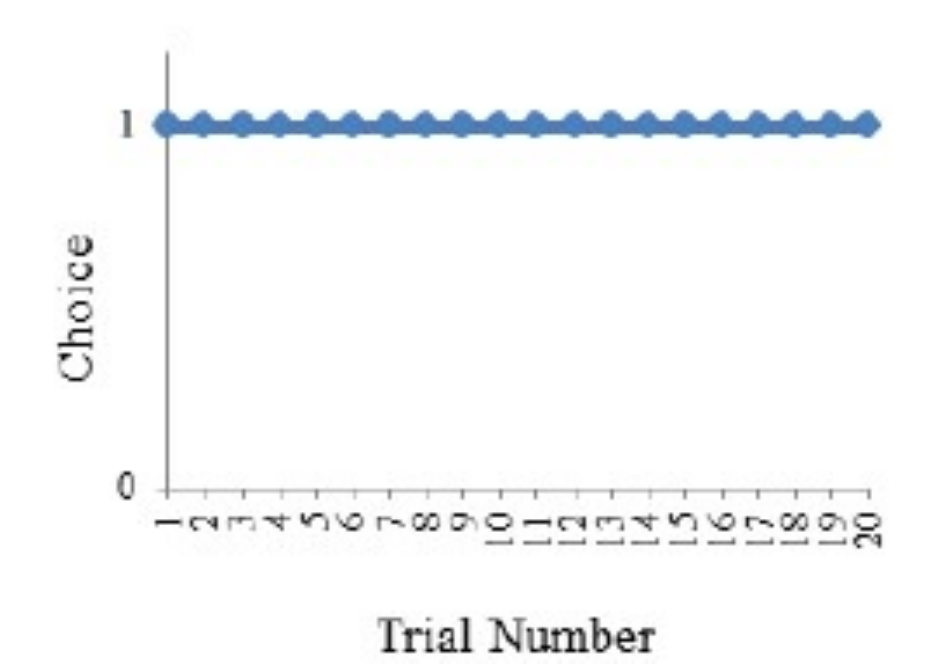}} & A driver starts by arbitrarily selecting a route, is apparently satisfied with the experience, and continues making the same choice for the entire 20 trials. \\ \midrule
  2& \raisebox{-3.0cm}{\includegraphics[scale=0.5]{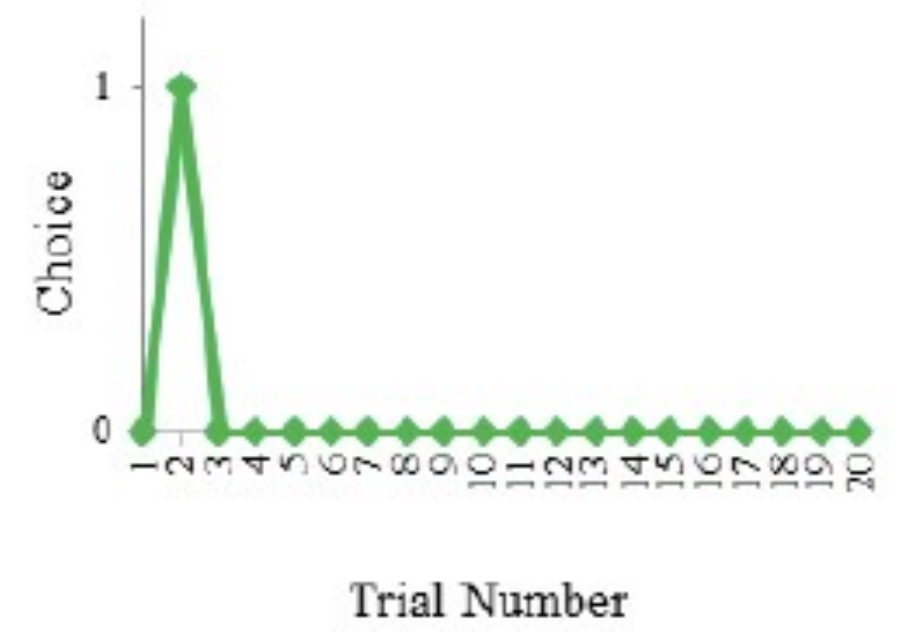}} & A driver starts by arbitrarily selecting a route, is apparently not satisfied with the experience, tries the other route, and decides that the first route was better. The driver makes a choice after trying both routes and does not change afterwards.  \\ \midrule
  3& \raisebox{-3.0cm}{\includegraphics[scale=0.5]{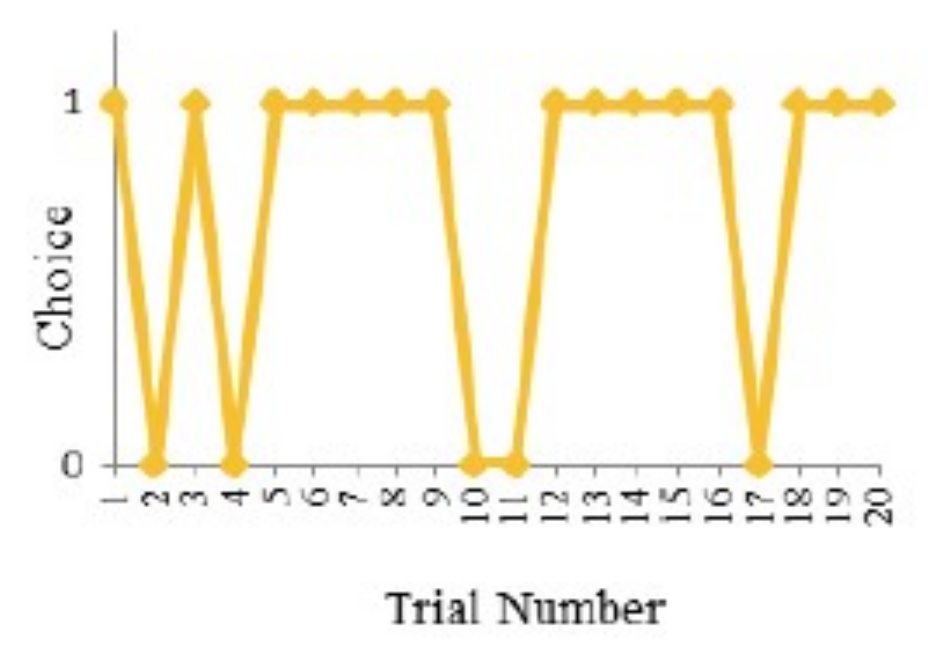}} & A driver switches between the two alternative routes over the duration of the experiment. The driver, however, drives on one route more than the other route. This reflects his/her preference for the selected route. \\\midrule 
  4& \raisebox{-3.0cm}{\includegraphics[scale=0.5]{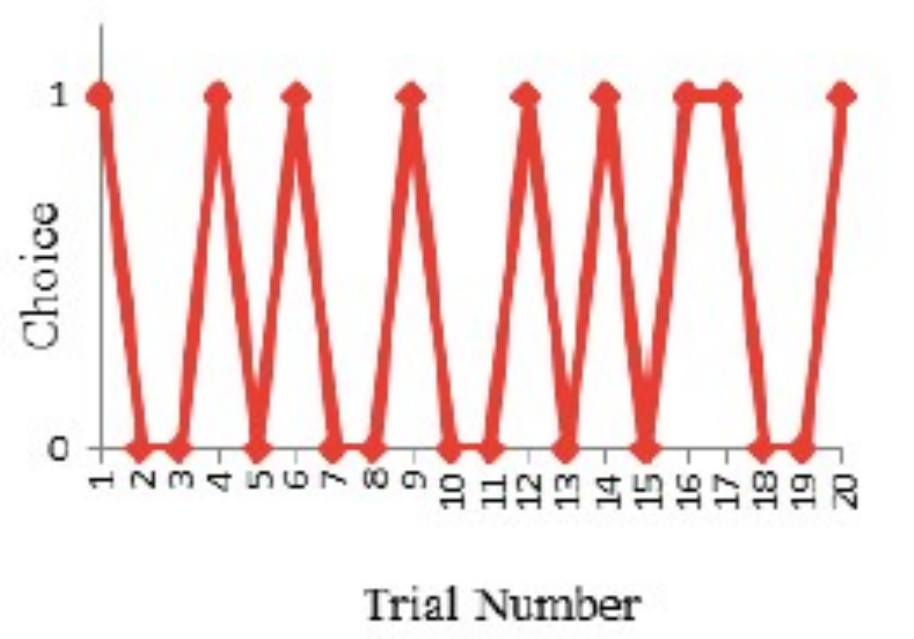}} & A driver switches between the two alternative routes over the duration of the experiment. The driver drives both routes with approximately equal percentages. This reflects a lack of preference towards any of the alternatives. \\\bottomrule[1.5pt]
\end{tabular}
\end{table*}
\section{Experimental Design}
As aforementioned, Tawfik et al. identified four route choice patterns observed in a real world experiment. This experiment attempts to quantify the influence of route information on traveler route choice behavior by comparing the choice patterns between Tawfik et. al.\textquoteright s experiment and the experiment conducted in this study. Occasionally, drivers prefer a route they frequently choose instead of switching to the actually faster route; or may deviate from the habitual route to the alternative route which is on average worse, only because the performance of the usually-taken route becomes bad on a random day. These irrational behaviors may probably be caused by a lack of precise information. The study attempts to address a number of questions: will travel time information make drivers behave more rationally? will the effect of information be different among individuals? what type of information will be most effective? \par 
            
A total of 20 participants were recruited within two age groups (18-33 and 55-75)\footnote[1]{These two age groups were selected because the authors wanted to investigate the impact of drivers’ age on the information effectiveness in changing choice behavior. The big difference of drivers’ age in the two age groups may more easily distinguish the difference of information effect attributed to drivers’ age.}, 10 male and 10 female. Each of them was required to accomplish three sectors of the experiment: a pre-run questionnaire, on-road test and a post-run questionnaire. The pre-run questionnaire was conducted before the beginning of the on-road test, which gathered the participants\textquoteright ~demographics, driving experiences, preferences, habits, information usage and the perception of route performance. Noticeably, each participant was demonstrated to have little knowledge of the route performance according to the results of pre-run questionnaire. The on-road test was conducted around the areas in Blacksburg and Christiansburg, VA for the morning, noon and evening peak from October 2013 to April 2014. The participants were asked to drive as if \footnote[2]{When drivers were doing the test, they were asked to drive from one predefined origin to the destination during every trip, and they actually did not commute during the test. But the researchers wanted to emulate the trip as a commute trip on which travel time may probably be the first consideration by the drivers.  So “as if” here means that drivers were asked to behave like commute.} they were commuting in order to ensure that travel time was an important consideration when they were to make choices. Each participant was asked to drive 11 trials, 5 of which provided participants with strict information (average travel time) and 5 provided with range information (travel time variability). The last one trial was not provided with any information, aiming to see how well information impacted drivers. It should be noted that the information was provided one time with average travel time and one time with travel time variability in order to eliminate the bias on each of the information type. The average travel time information provided to each trial was estimated by averaging the experienced travel time of three previous trials \footnote[3]{The experienced travel time was recorded by GPS during the testing; three previous trials were selected to be averaged because the trails before has little impact on the decision based on the literature \cite{tawfik2012real}; information used for the first trail was obtained from the experiment in \cite{tawfik2012real}} and travel time variability was estimated using the average value and standard deviation ($average\ travel\ time \pm 2*standard\ deviation$), so that the information could be dynamically updated each day to enhance the reliability estimate. It is worth noting that the provided travel time information was collected using GPS during the real-world experiment rather than from on-road or in-vehicle sensors, but the experimental design methodology of this study is also applicable to sensor-based information. For each trial, there were five O-D trips each of which had two alternative routes, one route was on average faster in travel time than the other. The characteristics of each route were specified in \autoref{tab.2}. The participants' task was to repeatedly make choices between the two alternatives on each trip. Statistically, 55 choice observations were collected for each participant, 100 observations by each trial and 220 on each trip. Upon the completion of 11 trials of the on-road test, the post-run questionnaire was thereafter conducted, whereby the participants were asked whether the provided information was beneficial. The accuracy of travel time perception would be compared between the two questionnaires in order to have a knowledge of whether the participants' perception was improved as a result of providing them with information.\par

The logical choice rate---the proportion of times in which the faster route is chosen as a function of time (trial number), participant and trip, respectively---was selected as the indicator of the positive role of information in facilitating rational behavior. The inertial choice rate---the proportion of participants remaining on their habitual but slower route---served to evaluate whether the information contributed to enhancing participant attitudes of risk seeking in the gain domain. The on-road data collected by Tawfik was applied to estimate the choice rates specified as \lq\lq without information\rq\rq ~group. Tawfik's experiment was conducted on the same trips in Blacksburg and Christiansburg in 2012, which was also a day-to-day commuting test in which participants were asked to repeatedly make choice between the two alternative routes on each trip. The difference between the two experiments was that the proposed study provided participants with travel time information.  For more details of Tawfik's study, see \cite{tawfik2012real}.\par
\begin{table*}
\centering
\caption{ROUTE CHARACTERISTICS OF EACH O-D TRIP}
\label{tab.2}
\captionsetup{font={scriptsize}}
\begin{tabular}{@{}ccm{2.3cm}cccm{4.2cm}@{}}\toprule[1.5pt]
\multirow{2}{1cm}{Trip No.}&\multirow{2}{1cm}{Route No.}&\multirow{2}{2.5cm}{Ave. Travel Time}&\multicolumn{2}{c}{No.of intersections}&\multirow{2}{2cm}{No.of left turns}&\multirow{2}{2cm}{Route description}\\
\cline{4-5}
& & & Signalized & Unsignalized & \multicolumn{2}{c}{}\\
\midrule[1.0pt]
\multirow{2}{*}[-0.1 cm]{1}&$1$&$9.2$&$10$&$3$&$3$& Mostly a high speed (65 mi/h) freeway\\
\cline{2-7}
&$2$&$9.3$&$5$&$4$&$4$& High speed (45 mi/h) urban highway\\
\hline
\multirow{2}{*}[-0.3 cm]{2}&$3$&$15.8$&$5$&$2$&$3$& Mostly a shorter, low speed (30 mi/h) back road with a lot of curves\\
\cline{2-7}
&$4$&$18.2$&$2$&$2$&$2$& Mostly a longer, high speed (55 mi/h) rural highway\\
\hline
\multirow{2}{*}[-0.2 cm]{3}&$5$&$8.6$&$5$&$3$&$3$& A longer high speed (65 mi/h) freeway followed by a low speed (25 mi/h) urban road\\
\cline{2-7}
&$6$&$9.4$&$8$&$3$&$2$& A shorter urban route (40 and 35 mi/h)\\
\hline
\multirow{2}{*}[-0.3cm]{4}&$7$&$10.4$&$5$&$3$&$4$& A short urban route that passes through campus (25 and 35 mi/h)\\
\cline{2-7}
&$8$&$10.3$&$6$&$2$&$2$& Primarily a long high speed (65 mi/h) freeway and low speed (25 mi/h) urban roads\\
\hline
\multirow{2}{*}[-0.3cm]{5}&$9$&$10.5$&$8$&$4$&$4$& A long urban road that passes through town (35 mi/h)\\
\cline{2-7}
&$10$&$8.5$&$3$&$1$&$3$& A short low speed (25 and 35 mi/h) rural road that passes by a small airport, and more direct\\
\bottomrule[1.5pt]
\end{tabular}
\end{table*}
\section{Results Analysis}
By comparing the perceived travel time of the pre-run questionnaire to the actual travel time collected during the on-road tests, it was demonstrated that the accuracy of participants' perception of travel time ranged from 5\% to 55\% for all five trips, with an average accuracy of only 38\%. Consequently, it would be safe to conclude that the participants had limited knowledge of the route performance prior to the start of the experiment. Based on the results of participants' perception in the post-run questionnaire, the average accuracy increased from 38\% to 62\% with an increase of 24\%. Consequently, it would be interesting to see whether participants behave more rationally with higher perception accuracy. \par

\begin{figure}
\centering
\includegraphics[width=8cm,height=8cm]{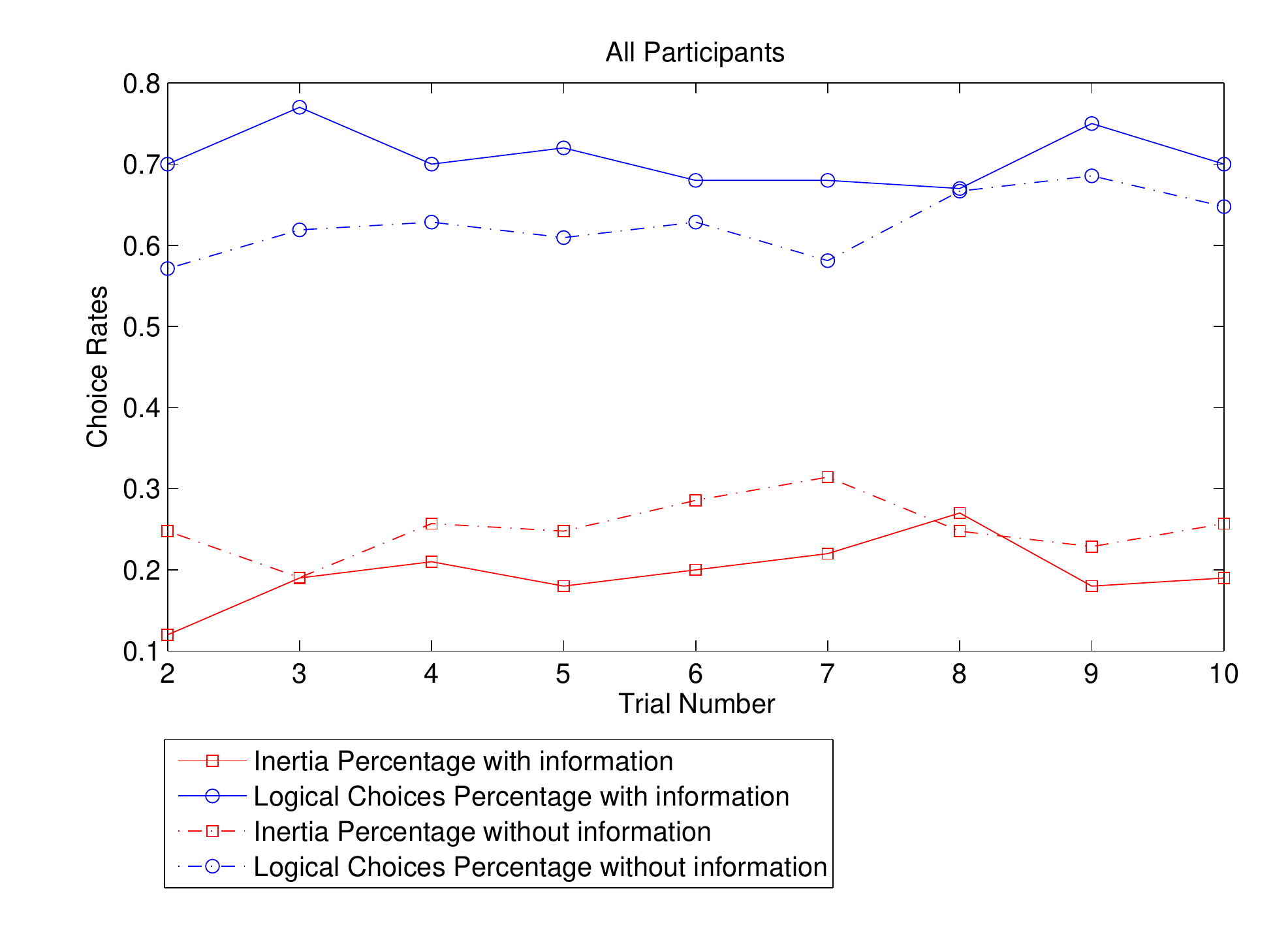}
\caption{Logical- and inertial- choice rates over trials.}
\label{Fig.1}
\end{figure}

\autoref{Fig.1} presents the proportions of logical- and inertial- choices as a function of time, identified as trial number. As expected, the logical choice rates are on average around 10\% higher in the \lq\lq with-information\rq\rq ~group than \lq\lq without-information\rq\rq ~group, especially for the first two trials in which the enhancement is up to 15\%. This demonstrates that the positive effect of information becomes more evident when travelers have limited knowledge of route performance. Although there are some oscillations at some of the trails, in general, the logical rates between the two groups are getting closer from the beginning to the end. The inertial choice rates are basically lower with the provision of information, implying that it is more likely for travelers to risk switching to the faster route when they are informed. However, regardless of being informed or not being informed, the inertial behavior is not reduced in day-to-day variation, which is different from the results in the simulation study \cite{{srinivasan2000modeling}}. This may be attributed to the habit or other decision considerations. \par
\begin{figure*}
\centering
\subfigure[Choice patterns without route information]{\label{fig:Participants choice patterns without vs. with route information:a}
\begin{minipage}[c]{1.0\textwidth}
\centering
  \includegraphics[width=1.0\textwidth]{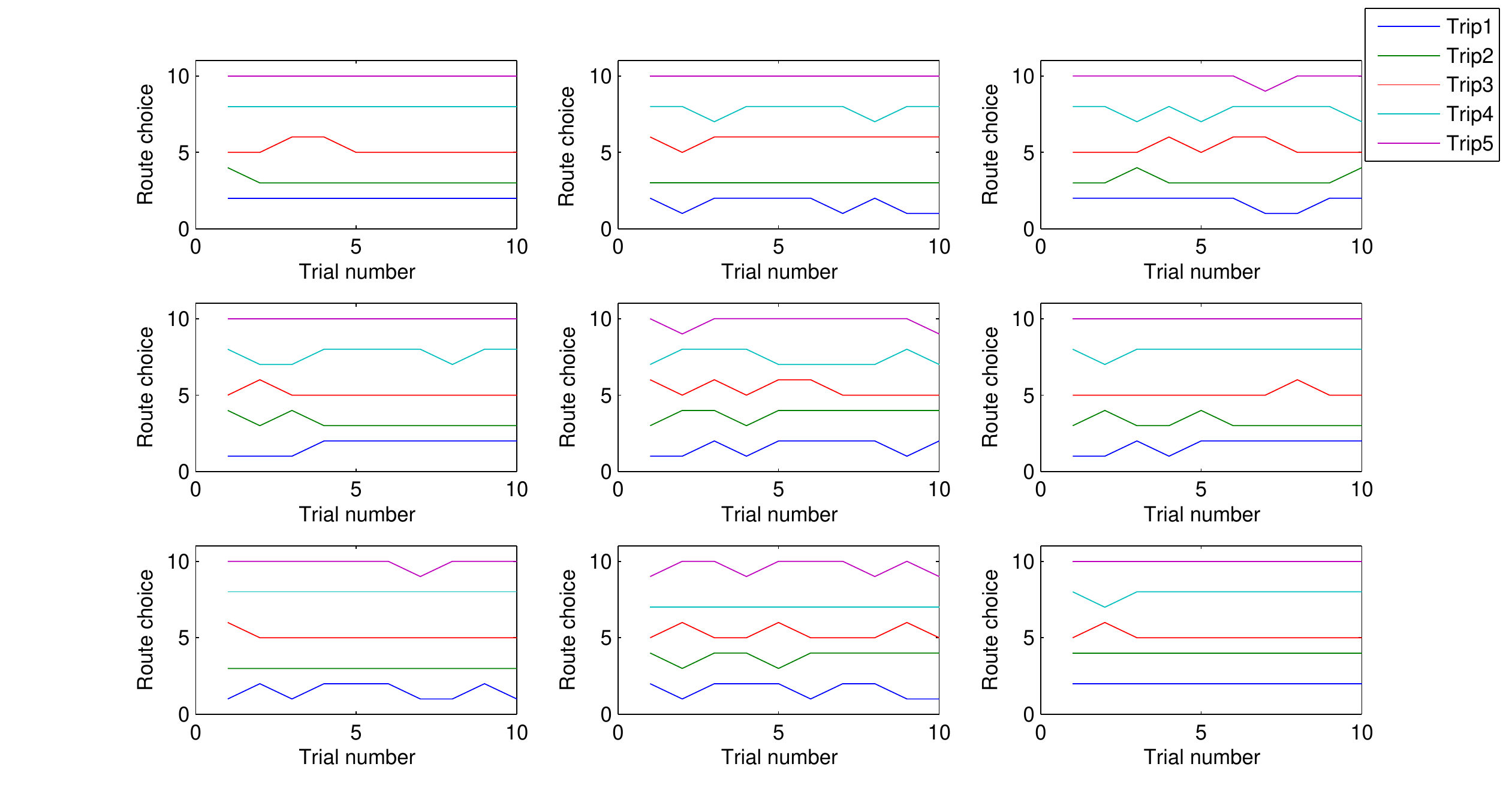}
\end{minipage}%
}
\subfigure[Choice patterns with route information]{\label{fig:Participants choice patterns without vs. with route information:b}
\begin{minipage}[c]{1.0\textwidth}
\centering
  \includegraphics[width=1.0\textwidth]{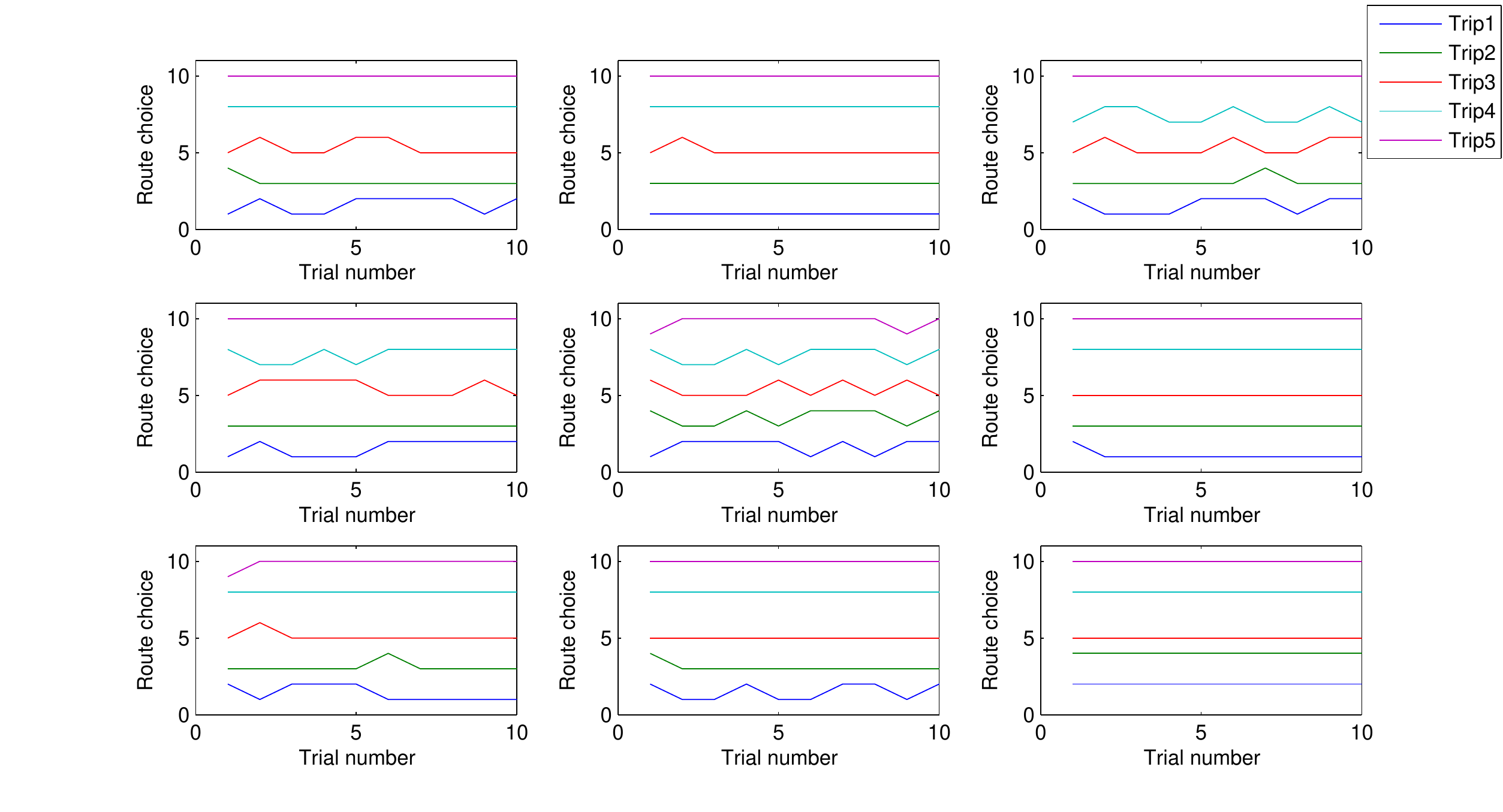}
\end{minipage} 
}
\caption{Participants choice patterns without vs. with route information.}\label{Fig.2}
\end{figure*}

In reality, the behavioral effect of information varies from person to person. One may probably have more confidence in his/her experiences than the acquired information; or travel time is not his/her top consideration. Accordingly, the insights gained from previous analyses are needed. Nine of the participants in this study attended Tawfik and Rakha's experiment. The choice results of these participants were specifically compared between the two experiments in order to see how the effect of information differentiated individually and how well they learned from the information. \autoref{Fig.2} compares the behavioral types (introduced in \autoref{tab.1} which was proposed by \cite{tawfik2013latent}) for each of the nine participants between with- and without-information. Only 10 trials were compared because the participants were informed for 10 trials only. The degree of the fluctuation of each line gives an explicit generalization of participants' behavioral aggressiveness. The more fluctuated in the lines, the more aggressively the participants behave. In general, the information significantly changes behavioral types either from risk-seeking to risk-aversion or vice versa. Some of the participants exhibited a high preference for one route when information was not provided and switched frequently when they were informed; whereas some switched more without information and maintained a single route when informed. Overall, the effect of information significantly differs at an individual level.\par
\begin{figure*}[h]
\subfigure[Logical choice rates]{
   \includegraphics[width=0.5\textwidth]{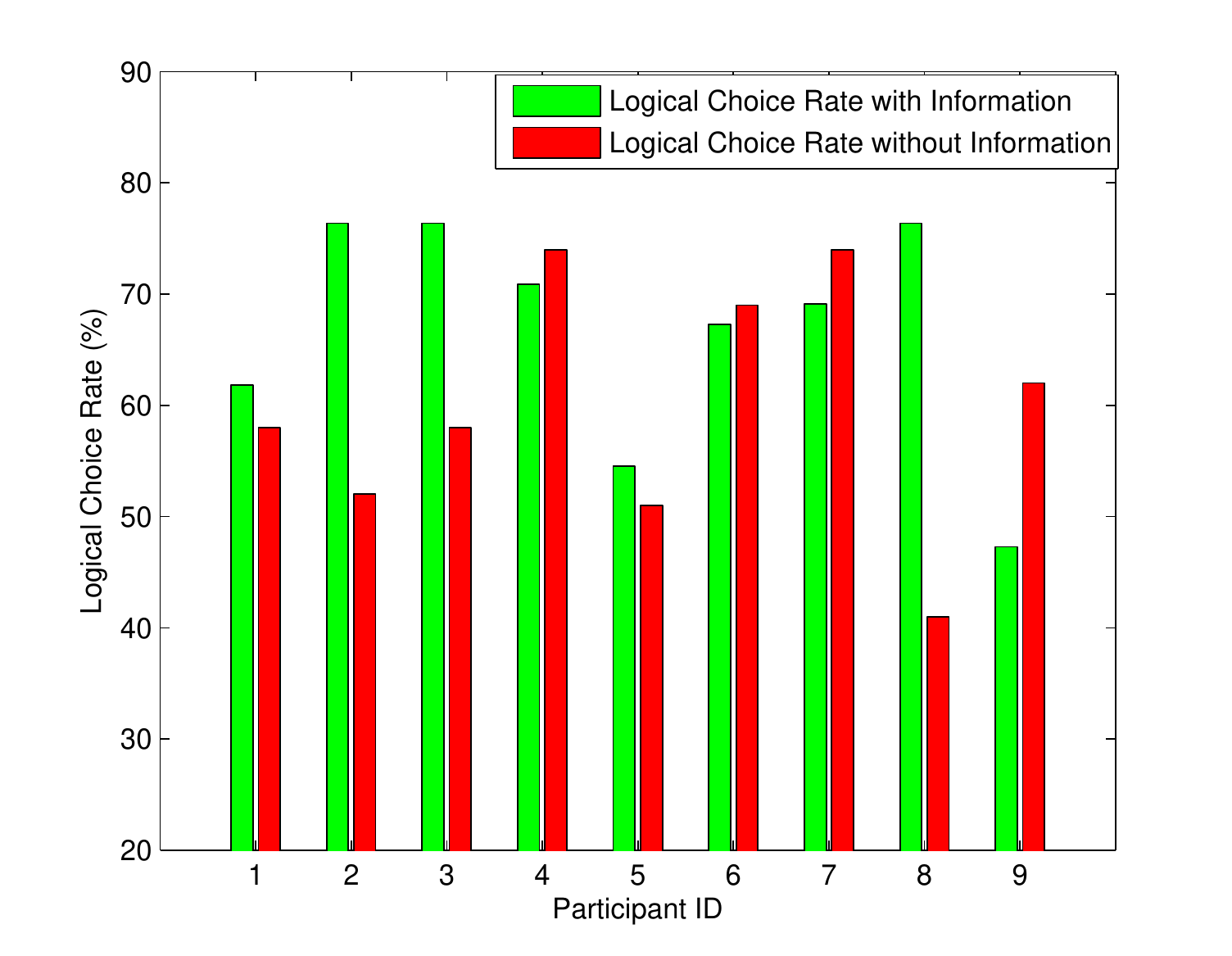}
 }
\subfigure[Inertial choice rates]{
   \includegraphics[width=0.5\textwidth]{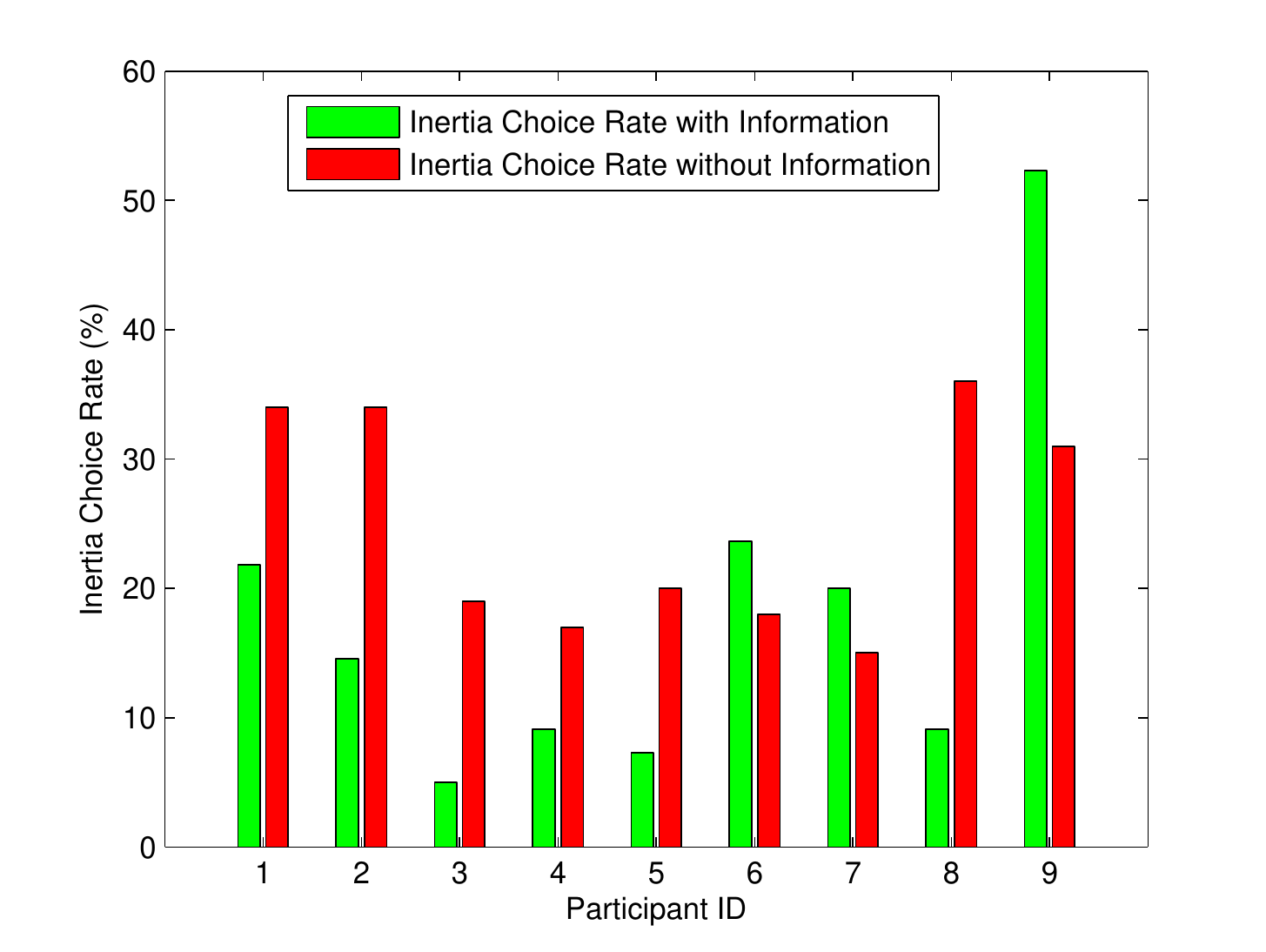}
 }
\caption{Logical- and inertial- choice rates over participants.}\label{Fig.3}
\end{figure*}

\autoref{Fig.3} summarizes the behavioral tendency of participants. According to \autoref{Fig.3}a, participants 1, 2, 3, 5, 8 basically moved their choices towards rationality with the assistance of information, whereas participants 4, 6, 7, 9 behaved more irrationally when they were informed. In \autoref{Fig.3}b, participants 6, 7, 9 instead have higher inertial rates with the provision of information, implying that they behaved even more risk aversive whey they were provided with information.\par
Based on the results of the post-run questionnaire, participants 6, 7 and 9 mentioned that travel time information had little impact on their route choices. Specifically, participant 6 preferred rural roads due to his preference on route scenery, although travel time was important to him as well. Participant 7 held the point that, instead of travel time, the number of intersections was the overriding factor she considered for route choice decisions. Participant 9 preferred to stick to her current route without any route-switching, which is the first type of the typical behavior shown in \autoref{tab.1}. Noticeably, participant 4 had both logical and inertial rates decreased with the provision of travel time information. That was because travel time was not the only consideration to this participant. Based on the results of the questionnaire, \lq\lq avoid traffic lights\rq\rq ~was the other equally important factor to him, which highly impacted his choice behavior. Occasionally, participant 4 switched to the slower route instead in order to avoid traffic lights even though he was informed the alternative route was better in terms of travel time, which increased the proportion of compromising behavior (the other type of illogical choice other than inertial choice) and decreased the logical choice rate. In general, travel time may have little effectiveness in enabling drivers to behave logically when drivers do not take travel time as their foremost factor in planning their routes. Additionally, participants 4, 6, 7, 9 are all senior persons from the age group of 55-75 year old. This implies that elder drivers are preferable to make choices based on their preferences or habits rather than received information, which confirms \cite{jou2005route}'s results.\par
\begin{figure*}
\subfigure[Logical choice rates]{
\includegraphics[width=0.5\textwidth]{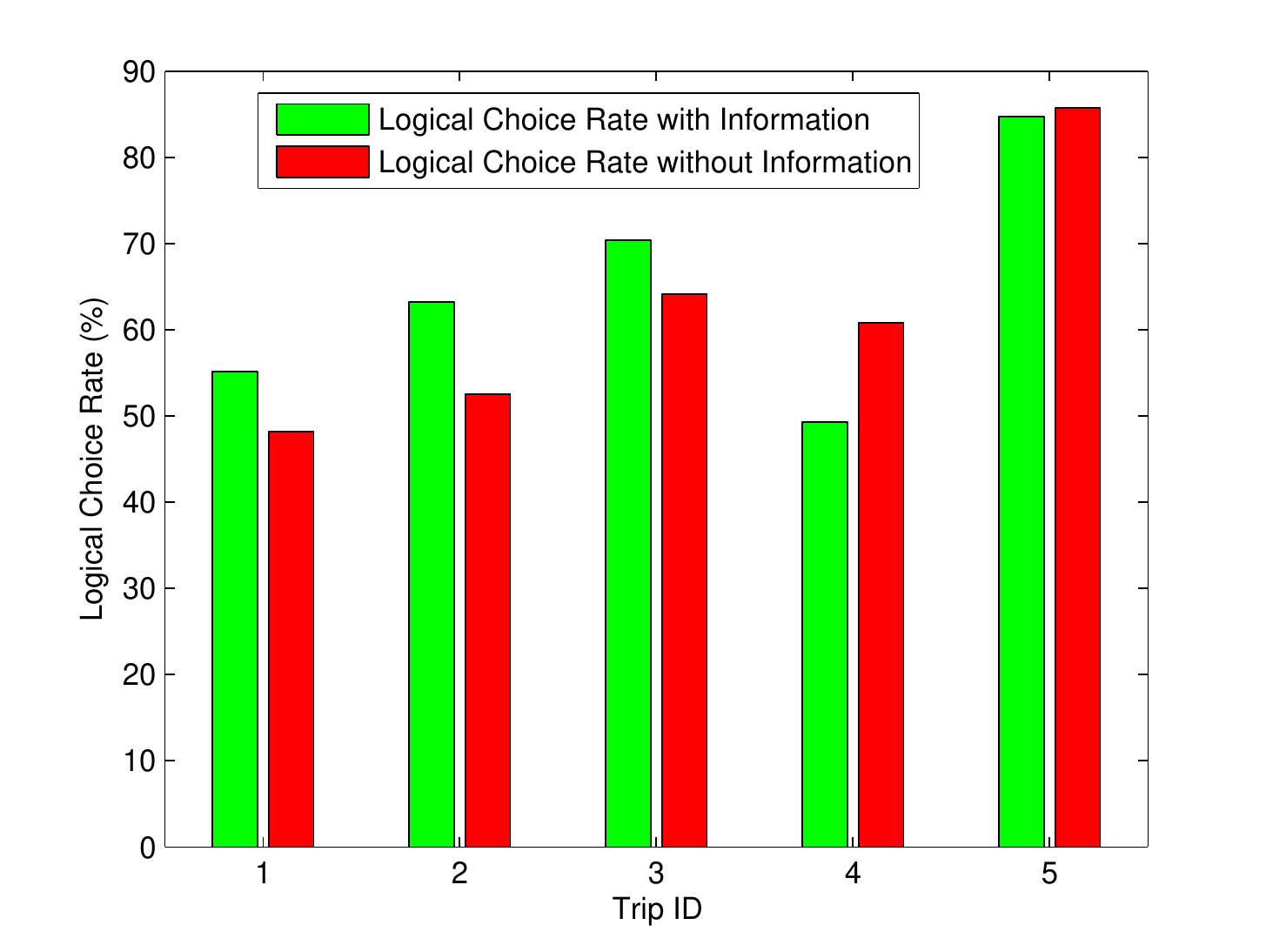}
}
\subfigure[Inertial- choice rates]{
\includegraphics[width=0.5\textwidth]{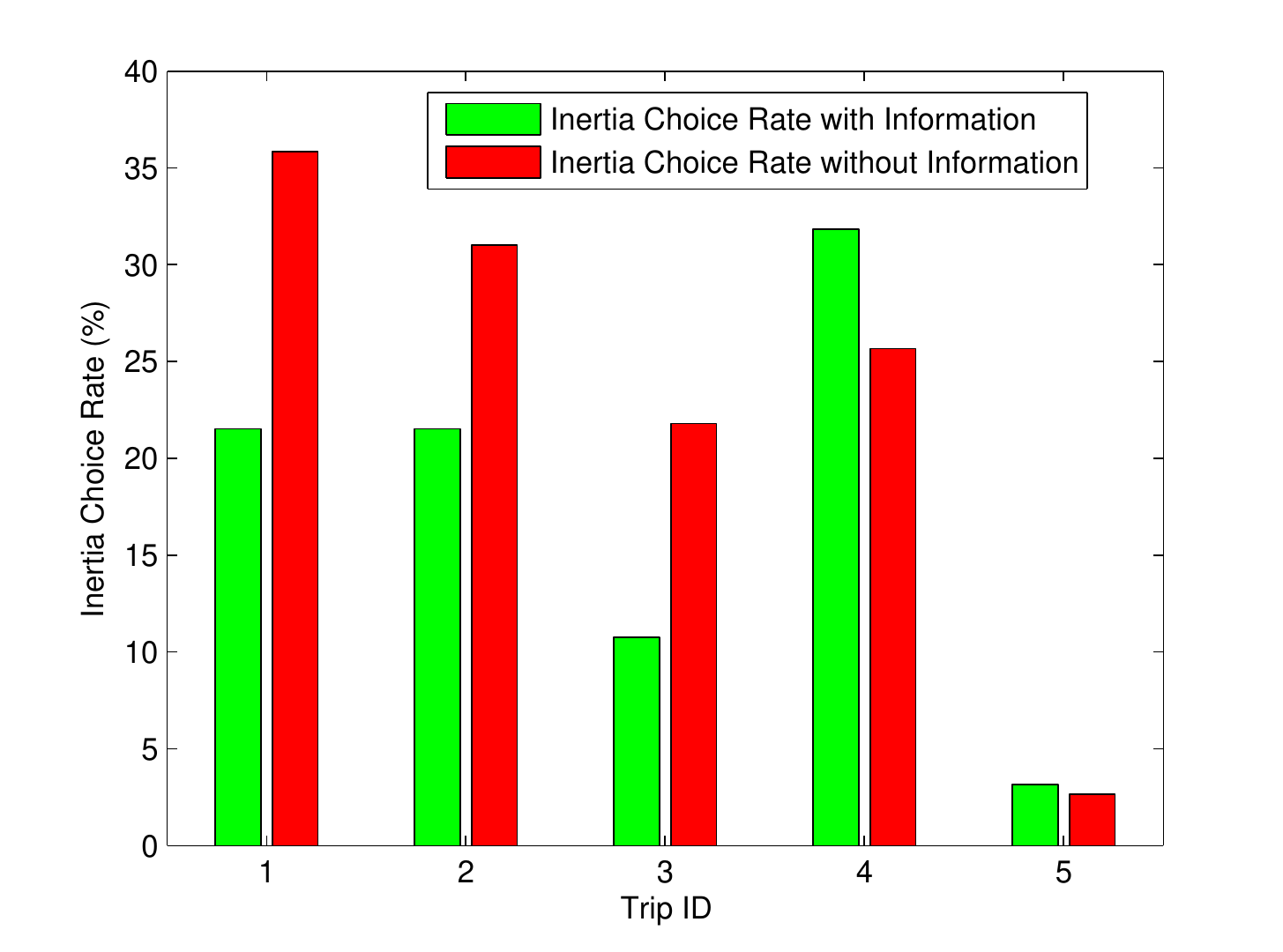}
}
\caption{Logical- and inertial- choice rates over trips}\label{Fig.4}
\end{figure*}

In addition to individual traits, trip characteristics may also affect the positive role of information. To study such effects, the choice rates were aggregated by trips. As illustrated in \autoref{Fig.4},  information enhances behavioral rationality only for the first three trips. For trip 4, logical rates decrease while inertial rates increase when information is provided. On trip 5, the choice rates do not change significantly between with- and without- information. According to the route characteristics addressed in \autoref{tab.2}, route 7 and route 8 (on trip 4) are almost identical in travel time, whereas many participants pointed out that they were reluctant to take route 7 even though it occasionally took less travel time since they did not want to risk being caught on campus by pedestrian flows. The provided information was considered to be less reliable for this trip. Interestingly, travel time is very close as well between the two routes on trip 1; however, the effect of information appears to be very positive. That is because there is no distinct advantage for one route over the other on this trip. Although route 1 is on a highway system with a 20 km/h higher speed limit than route 2, there are five more signalized intersections on it. The provided travel time information for this trip was considered reliable by participants. For trip 5, route 10 distinctively outperforms route 9 in terms of travel time, directness, less traffic and fewer intersections. \cite{tawfik2012real} clearly indicated that drivers were able to precisely perceive the route performance and to make correct decisions on this trip without any assistance of information. Overall, information provides little benefit if one route visibly outperforms the other.\par
\begin{figure*}
\subfigure[Choice rates over trials]{
   \includegraphics[height=0.5\textwidth,width=0.5\textwidth]{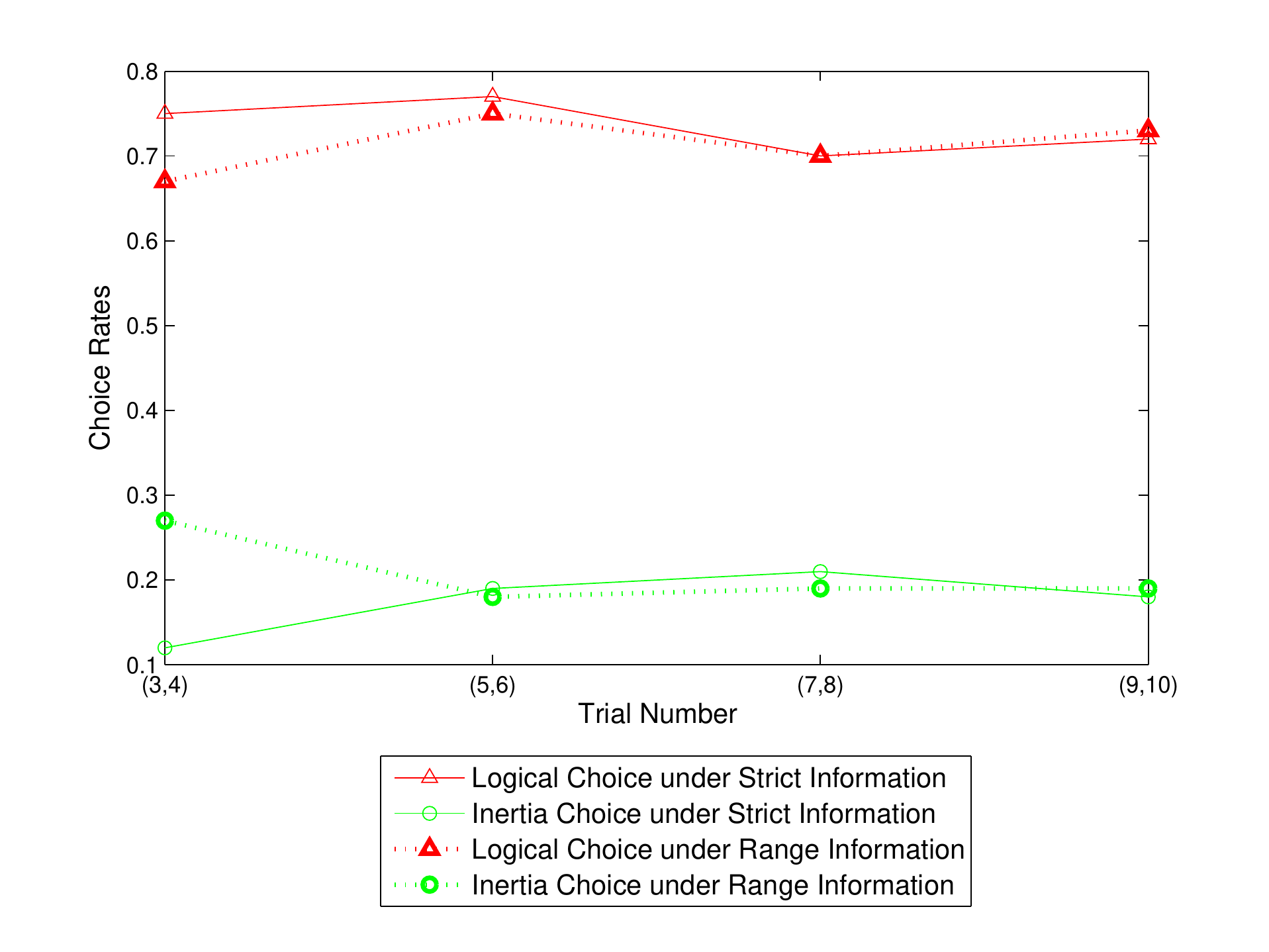}
 }
\subfigure[Choice rates over participants]{
   \includegraphics[height=0.5\textwidth,width=0.5\textwidth]{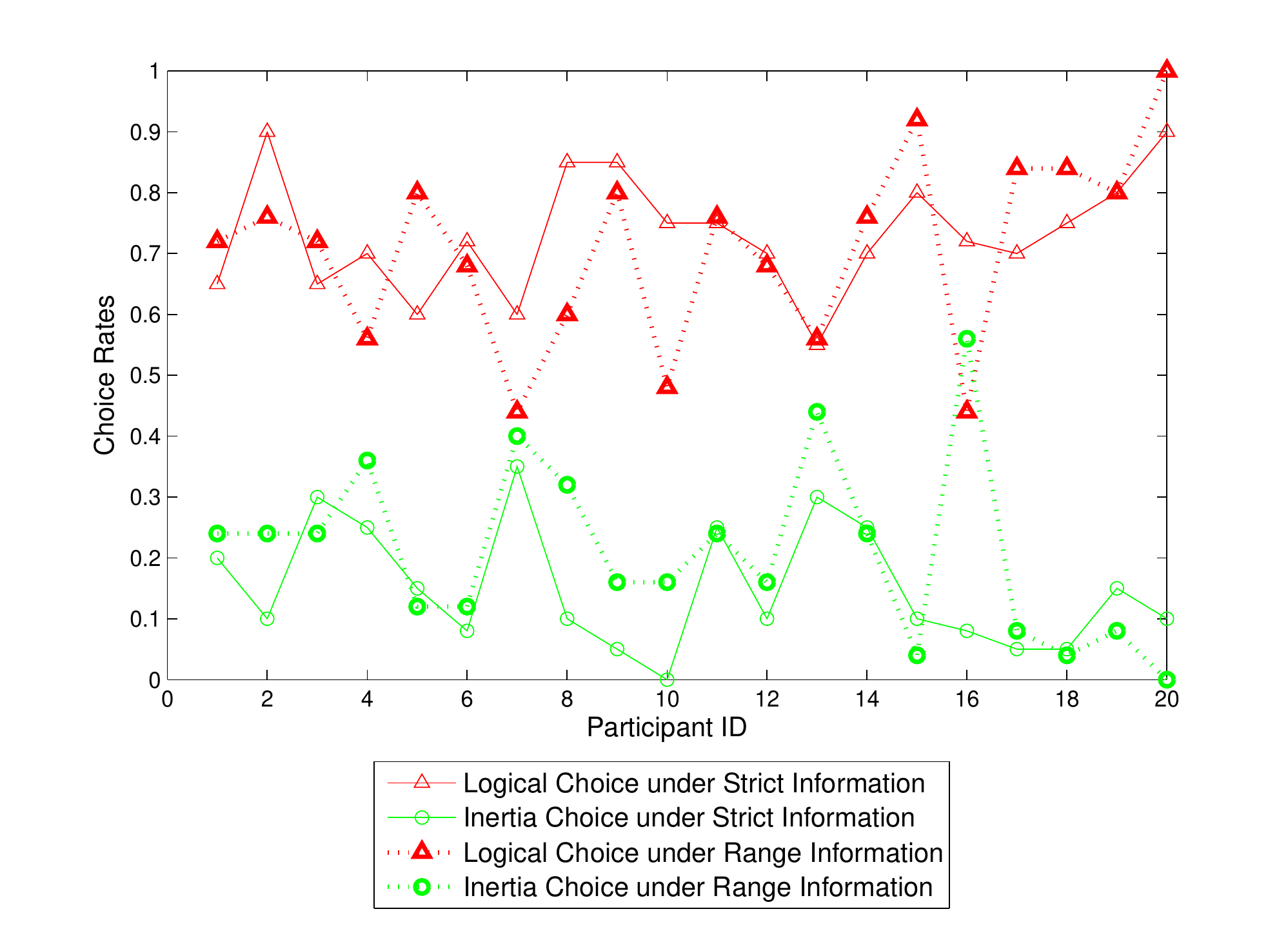}
 }
\caption{Choice rates with strict information vs. with range information}\label{Fig.5}
\end{figure*}

\begin{figure*}
\subfigure[Choice rates over trials]{
   \includegraphics[height=0.5\textwidth,width=0.5\textwidth]{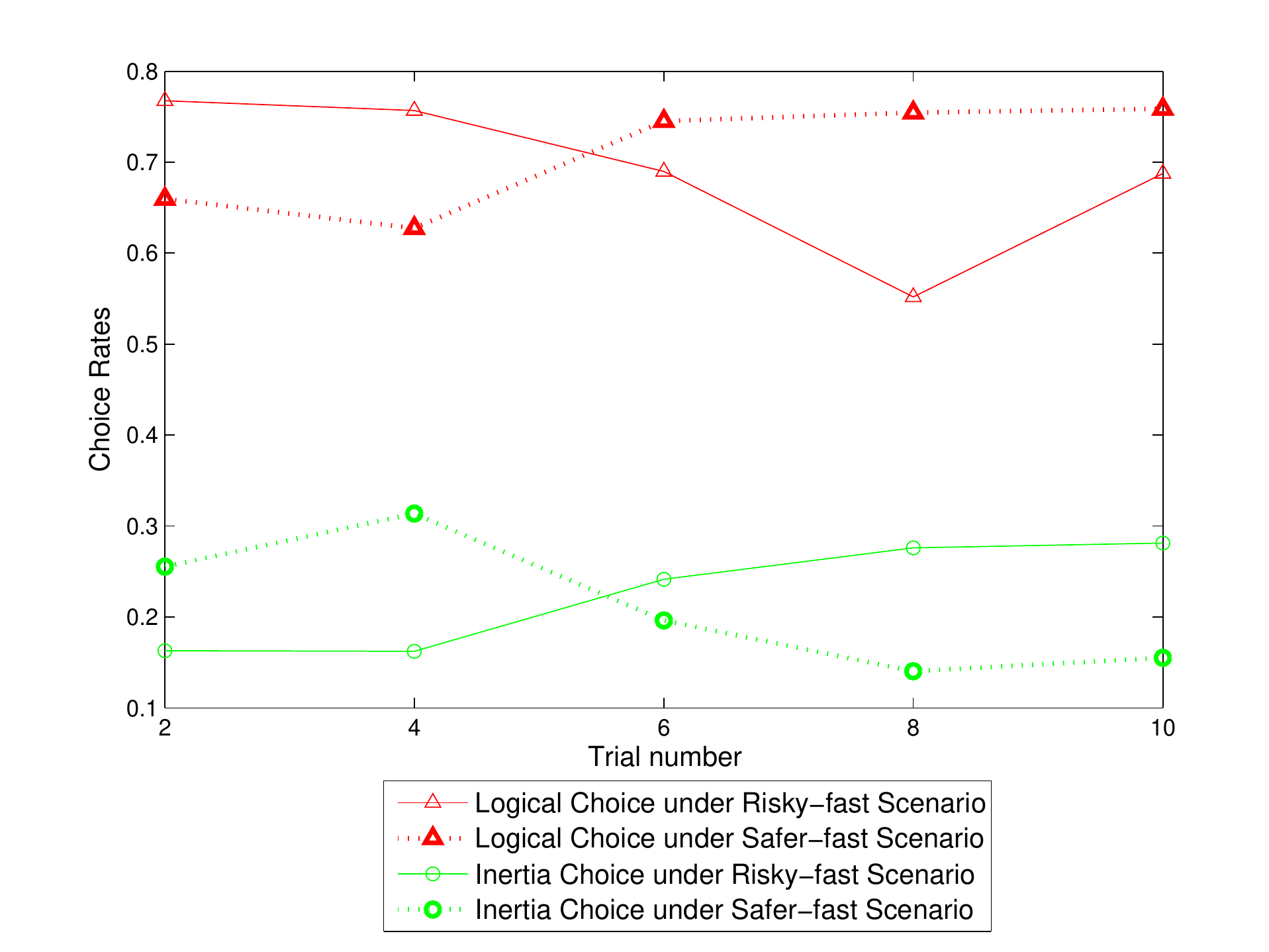}
 }
\subfigure[Choice rates over participants]{
   \includegraphics[height=0.5\textwidth,width=0.5\textwidth]{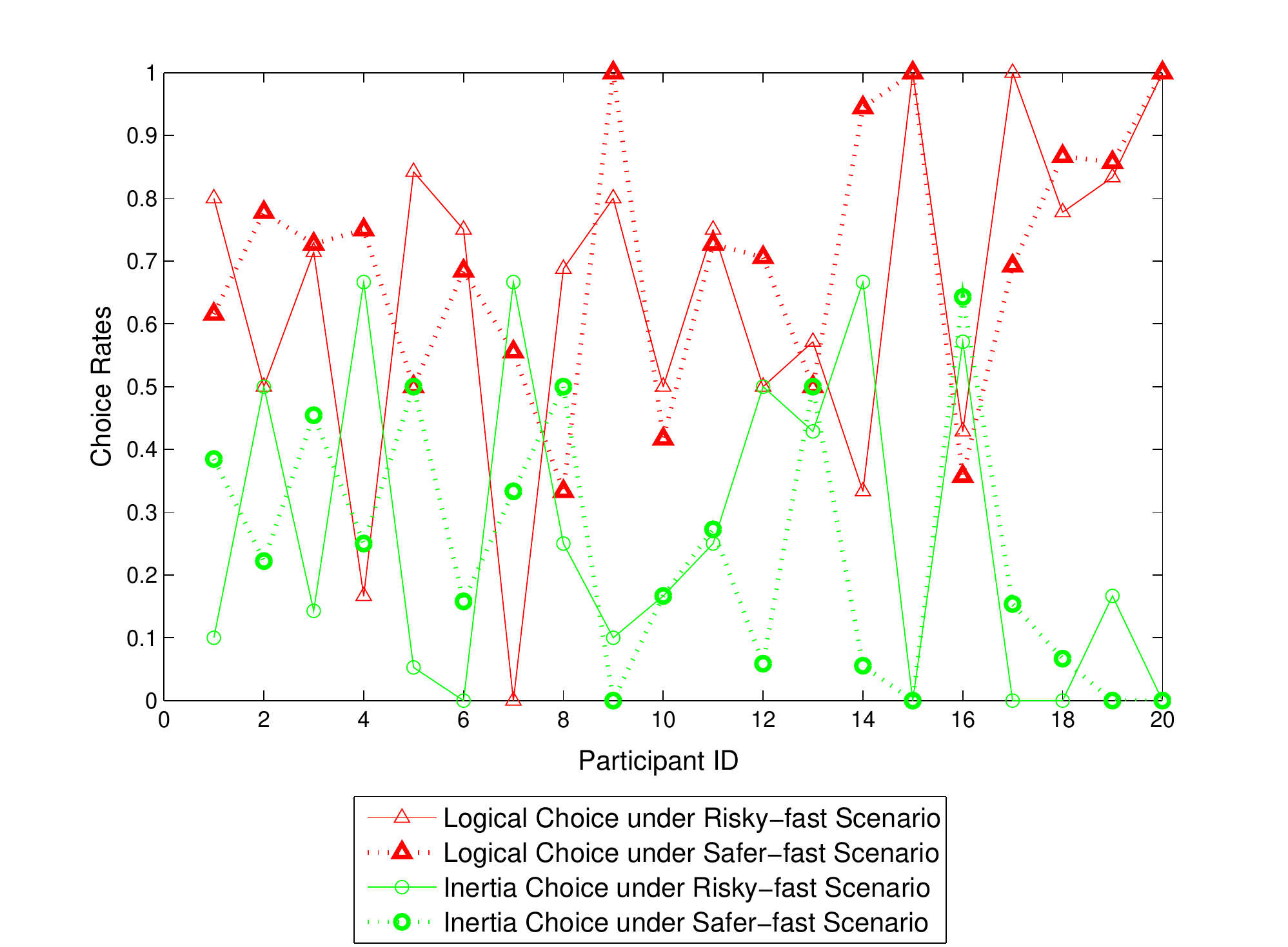}
 }
\caption{Choice rates with risky-fast scenario vs. safer-fast scenario}\label{Fig.6}
\end{figure*}

\autoref{Fig.5} and \autoref{Fig.6} provide a broad view of the effect of different information types on route choice behavior. In \autoref{Fig.5}, the comparative analysis was performed between strict information (average travel time) and range information (variability). According to \autoref{Fig.5}a, strict information results in higher logical rates with lower inertial rates for the first trial, demonstrating that strict information is more effective than range information when drivers are lack of experience. For the following trials, however, there is no significant distinctiveness between the two scenarios. This may be attributed to the fact that the effect of information type tends to be identical after drivers gain experience. As illustrated in \autoref{Fig.5}b, strict information results in higher logical rates and lower inertial rates on average. Nonetheless, to some of the participants, range information performs better, implying that the responses to different information types, to a large extent, are dependent on individual traits, although strict information overall performs better in this study. \par

\autoref{Fig.6} presents the effect of different range information scenarios. As illustrated in \autoref{Fig.6}a, “Risky-fast” scenario refers to the faster route (lower average travel time) with higher variability while “safer-fast” represents the faster route with lower variability. Interestingly, the risky-fast scenario appears to have higher logical rates and lower inertial rates in the first two trials; whereas the positive effect decreases in the following three trials. This implies that, when drivers have limited knowledge of route performance, the faster route with high variability is more attractive and subject to make drivers take risk in the gain domain. Once drivers gather experience, however, they are reluctant to risk seeking in the gain domain under higher uncertainty; instead, the safer-fast route becomes preferable. This confirms the result of \cite{katsikopoulos2002risk, ben2008combined,ben2010road, katsikopoulos2000framing}. \autoref{Fig.6}b demonstrates that there is no consensus between participants on which scenario is more effective. Some of the participants have higher logical rates and lower inertial rates for the risky-fast scenario while some exhibit the opposite pattern. \par
\section{Conclusions}
This study empirically investigates the effect of dynamic travel time information on day-to-day commuter route choice behavior by designing and running a real world experiment. The experiment confirms some of the results obtained from previous simulation studies, demonstrating that, in general, real-time information significantly enhances behavioral rationality especially when drivers lack long-term experience. Simultaneously, inertial choice rates decrease with information provision, demonstrating that drivers are more willing to risk switching to faster routes when they have more information about these routes. Nonetheless, the positive role of information is, to a large extent, dependent upon the individual's age, preferences, and route characteristics. The results demonstrate that travel time information may not have positive impacts on driver route choice behavior if they value other factors in making their decisions, such as route scenery, habit, number of intersections and traffic signals. The results also reveal that the effect of information on driver behavior is less evident for elder drivers, which is consistent with \cite{jou2005route}. In addition to personal traits, route characteristics are found to be another important factor influencing the effectiveness of information. Specifically, information may not add value if one route is significantly better than the other given that drivers would be able to identify the optimum route on their own through their experiences.\par 

The effect of the type of route information provided to the travelers on their route choice behavior was also studied. The conclusions are consistent with the results of simulation studies, demonstrating that, when drivers have limited experiences, information on expected travel times is in general more effective than information on travel time variability in enhancing rational behavior. After drivers gain sufficient knowledge of the alternative routes, however, the benefit of providing strict information appears to diminish. The results also demonstrate that drivers prefer to take the faster less reliable route as opposed to the slower more reliable route when they lack historical experience. However, as drivers accumulate experience, they become more willing to take the more reliable route, demonstrating that they become less risk seeking in the gain domain at higher uncertainty once experience is gained. In addition, the effect of information types on route choice behavior significantly differs from person to person. Which type of information is most effective to what group of travelers remains to be investigated in future research.\par

The experiment also demonstrates that, regardless of being informed or not being informed, the drivers' inertial behavior does not reduce in day-to-day variation, which is different from the results obtained by the simulation study \cite{{srinivasan2000modeling}}. This may be attributed to the habit or much more decision considerations in actual driving conditions. 

Finally, it should be noted that given the small sample size, these conclusions serve as a first attempt at understanding driver route choice behavior empirically. Further research is needed to validate these findings on a bigger sample of drivers and for different confounding factors.
\section{Acknowledgements}
 This effort was funded by the Mid-Atlantic University Transportation Center (MAUTC). The authors acknowledge all personnel who assisted with the data collection.
\section{References}
  \bibliographystyle{elsarticle-harv}
  \bibliography{mybib}
\end{document}